\documentclass[twoside,fleqn]{article}
\usepackage{espcrc2}
\usepackage[dvips]{epsfig}
\title{
{
\vspace{-3.0cm} \normalsize \hfill
\parbox{30mm}{SFB/CPP-03-33\\ HU-EP-03162}
}\\[15mm]
A precise determination of the decay constant of the D$_{\rm{s}}$-meson in quenched QCD
\thanks{Talk presented by Andreas J\"uttner at the International Symposium on Lattice Field Theory, Ibaraki, Japan, July 15-19, 2003. Work supported by the DFG in the SFB/TR 09 and in the GK271. We thank the DESY and the HLRN \cite{HLRN} for allocating computer time to this project.}
}
\author{{Andreas J\"uttner and Juri Rolf (ALPHA collaboration) }
\address{Humboldt-Universit\"at zu Berlin, 
              Institut f\"ur Physik,
              Newtonstr. 15,
              12489 Berlin}
}
\begin{document}
\begin{abstract}
We present a precise determination of the leptonic decay constant $F_{\rm{D_s}}$ of the ${\rm{D_s}}$-meson in quenched lattice QCD. This work is particularly focused on the analysis and discussion of all sources of systematic errors. We simulate directly at the physical quark masses for five different lattice spacings between $0.1\,$fm and $0.03\,$fm using $O(a)$-improvement. The finest lattice is still work in progress. After taking the continuum limit and setting the scale with the Kaon decay constant $F_{\rm K}=160\,$MeV we arrive at a value of $F_{\rm{D_s}}=252(9)\,$MeV. Setting the scale with the nucleon mass instead leads to a decrease of about $20\,$MeV of $F_{\rm{D_s}}$.
\end{abstract}
\maketitle
\section{Introduction}
The ${\rm D_s}$ meson consists of a charm and a strange quark. Its direct simulation on the lattice is possible without having to use a chiral extrapolation for the light quark or an effective theory for the heavy quark (e.g. HQET, NRQCD). 
Therefore the ${\rm D_s}$ is an ideal testing ground as its properties are expected to be similar to those of other heavy-light systems (e.g. ${\rm B_{(s)}}$).

The decay constant $F_{\rm{D_s}}$ parameterizes the leptonic decay of the ${\rm D_s}$ meson into a pair of leptons with a $W$ as the intermediate vector boson. It is defined through the QCD matrix element
\begin{equation}
\langle0|A_\mu(0)|{{\rm D}}_s(p)\rangle=ip_\mu F_{\rm{D_s}},
\end{equation}
where $A_\mu=\bar{\rm s}\gamma_\mu\gamma_5 {\rm c}$ is the axial current.

$F_{\rm{D_s}}$ is an important input parameter to CKM-physics in phenomenology. Together with a precise computation of the decay constant in the static approximation \cite{DellaMorte:2003mn,Micheletalk}, it can also be used to investigate down to which heavy quark mass HQET can be applied safely. With this at hand one can then eventually get a reliable estimate for $F_{\rm B}$ from an interpolation between the relativistic simulation and the static approximation \cite{Juritalk}.

Previous quenched results for $F_{\rm{D_s}}$ have been summarized in \cite{Ryan:2001ej}, which gives a world average of $230(14)\,$MeV. Ref. \cite{deDivitiis:2003wy} is a recent calculation of $F_{\rm{D_s}}$ with an alternative approach which gives a value of $F_{\rm{D_s}}=239(10)\,$MeV. The PDG quotes $F_{\rm{D_s}} = 285\pm   19 \pm  40\,$MeV \cite{PDG}. 

The aim of this project is to obtain a precise value for the decay constant $F_{\rm{D_s}}$ with a combined statistical and systematic error of 3\%, apart from the quenched approximation. 

The systematic errors can be kept under control by using non-perturbative $O(a)$-improvement \cite{Luscher:1997ug,Bhattacharya:2001ks} and renormalization \cite{Luscher:1997jn} together with a well controlled approach to the continuum. In particular, we simulated at four different values of the lattice spacing between $a=0.1,\dots,0.05\,$fm. At the moment we simulate with an even finer lattice ($0.03\,$fm) and we plan to add the results to the analysis in the future. We also analyzed the systematic error due to the limited solver precision and the systematic error due to the contamination of the plateaus by excited states and glueballs. Finite volume effects have been shown to be negligible at our choice of the lattice size and at our choices of the quark masses \cite{Garden:1999fg}.
\section{Strategy and Simulation} 
Our simulation parameters are based on previous work of the ALPHA-collaboration.
We use the Sommer scale \cite{Guagnelli:1998ud} with $r_0=0.5\,$fm to fix the lattice spacing. As shown in \cite{Garden:1999fg}, this choice of $r_0$ is equivalent to setting the scale with the decay constant of the K-meson, $F_{\rm K}=160\,$MeV. 
The hopping parameters of the strange quark and the charm quark have been taken from publications \cite{Garden:1999fg,Rolf:2002gu} on the corresponding renormalization group invariant masses for all values of the lattice spacing. In these works, $\kappa_{\rm s}$ and $\kappa_{\rm c}$ were determined using the masses $r_0m_{{\rm D_s}}=4.99$ and $r_0m_{\rm K}=1.57$ of the ${\rm D_s}$- and K-meson as phenomenological input, while neglecting isospin breaking and using $M_{\rm s}/M_{\rm light}=24.4\pm1.5$ with $M_{\rm light}=\frac{1}{2}(M_{\rm u}+M_{\rm{d}})$ from chiral perturbation theory \cite{Leutwyler:1996qg}.
\begin{table}[t]
\centering
\begin{tabular}{lllll}
\multicolumn{5}{l}{\hspace{-.8cm}Table1. Simulation data}\\
\hline
      $\beta$ & $n_{\rm{meas}}$   & $L/a$ & $L/r_0$ &  $r_0 m_{{\rm D_s}}$ \\ \hline
      6.0     & 380               & 16    &   2.98  & 4.972(22)     \\
      6.1     & 301               & 24    &   3.79  & 4.981(23)     \\
      6.2     & 251               & 24    &   3.26  & 5.000(25)     \\
      6.45    & 289               & 32    &   3.06  & 5.042(29)     \\
\hline
\end{tabular}
\label{simdata}
\vspace{-.7cm}
\end{table}

All calculations have been per\-for\-med in quenched QCD using clover-improved Wilson-Fermions and Schr\"odinger Functional \cite{Luscher:1992an} boundary conditions. In the time directions all fields obey Dirichlet boundary conditions and in the space directions they obey periodic boundary conditions (up to a phase in the fermion fields, which we chose to be $\theta=0.5$).  The fermionic boundary fields $\zeta,\bar{\zeta}$
 and $\zeta^\prime,\bar{\zeta}^\prime$
 are used to construct the mesonic boundary sources \cite{Luscher:1996sc}
\begin{equation}
\begin{array}{lcll}
\mathcal{O}&=&\frac{a^6}{L^3}\sum\limits_{\vec{y},\vec{z}}\bar{\zeta}_s(\vec{y})\gamma_5\zeta_c(\vec{z}),& {\rm at }\;\;x_0=0,\\
\mathcal{O^\prime}&=&\frac{a^6}{L^3}\sum\limits_{\vec{y},\vec{z}}\bar{\zeta}_s(\vec{y})\gamma_5\zeta_c(\vec{z}),& {\rm at }\;\;x_0=T.
\end{array}
\end{equation}
Starting from the correlation function 
\begin{equation} 
f_A^I(x_0)= -\frac{L^3}{2}\langle A_0^I(x) \mathcal{O}\rangle,
\end{equation}
with the improved axial current $A_\mu^I$, one derives the expression
\begin{equation}\label{fdsformula}
\begin{array}{lcl}
F_{\rm {D_s}}(x_0)&\approx&-2Z_A(1+b_A am_q)({ m_{\rm D_s}}L^3)^{-1/2}\\
&&\times\frac{ {f_A^I}}{\sqrt{{ f_1}}}e^{(x_0-T/2){ m_{\rm D_s}}}\\
&&\times\left\{1-\eta_A^{\rm D_s}e^{-x_0\Delta}-\eta^0_Ae^{-(T-x_0)m_G}\right\}\\
&&+O(a^2)
\end{array}
\end{equation}
for the decay constant. A detailed discussion can be found in \cite{Juttner:2003ns}.
\begin{figure}[t]
\vspace{-.9cm}
\caption{The simulation result for the bare decay constant for $\beta=6.0,6.1,6.2,6.45$ (from top to bottom). The filled symbols mark the a priori chosen range of the plateau.}
\vspace{-1.cm}
  \centerline{\epsfig{file=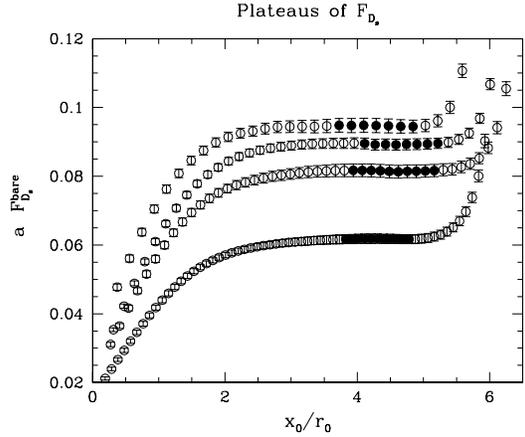,width=\linewidth}}
\label{fdsbare}
\vspace{-1.3cm}
\end{figure}
Here, $Z_A$ renormalizes the axial current and the term linear in $b_{\rm A}$ subtracts $O(a)$-artefacts proportional to the bare subtracted quark mass $m_q$. 
Both coefficients were obtained non-perturbatively on the lattice \cite{Luscher:1997jn,Bhattacharya:2001ks}. The pseudoscalar mass $m_{\rm D_s}$ is obtained as effective mass and $({ m_{\rm D_s}}L^3)^{-1/2}$ normalizes one particle states.
$f_A^I$ and $f_1=-\frac{1}{2}\langle\mathcal{O}^\prime\mathcal{O}\rangle$, which cancels out the normalization on the boundary meson sources, have to be determined in the simulation.
As $f_A^I$ decays exponentially with the pseudoscalar mass, one expects $F_{\rm {D_s}}(x_0)$ to exhibit a plateau at intermediate times, where the contributions $\eta_A^{{\rm D_s}} e^{-x_0\Delta}$ of the first excited state and the contribution $\eta_A^0 e^{-(T-x_0) m_{\rm{G}}}$ from the $O^{++}$ glueball both are small.  A plateau average can then be performed to increase the signal.

Some simulation parameters are summarized in table 1. We set $T=2L$. All simulations were done at approximately constant physics, which is important for the scaling study. 
Fig. 1 shows the simulation result for the unrenoramlized $F_{\rm D_s}(x_0)$.
\section{Error analysis and results}
The simulations for the first four values of $\beta$ were done in single precision arithmetics on the APE-machine at DESY Zeuthen. As we are dealing with heavy quark propagators, we have to check for rounding errors which may arise during the calculation.
We also want to guarantee sufficient solver convergence for the propagators. The corresponding checks were done with an adapted version of the MILC code \cite{MILC} on the IBM p690 computers at the HLRN \cite{HLRN}. Starting from the same field configuration, we calculated the propagators in single and double precision arithmetics and with different solver residuals. It turned out that the impact of these changes on $F_{\rm{D_s}}$ is below 1 per mil.

For the definition of plateau ranges of (\ref{fdsformula}) where the relative contribution of excited states and glueballs is below a threshold of 5 per mil, 
one needs estimates of $\Delta$ and $m_G$. 
These were obtained self-consistently from linear fits to $\log(F_{\rm D_s}^{\rm bare}(x_0)-F_{\rm D_s}^{\rm plateau})$, where $F_{\rm D_s}^{\rm plateau}$ is the average of an a priori chosen plateau (cf. fig. (\ref{fdsbare}) and \cite{Juttner:2003ns} for details).
We found that a suitable range for the plateau is $x_0=4r_0,\dots,5r_0$ for all values of $\beta$. 

\begin{figure}[t]
\vspace{-.9cm}
\caption{Continuum extrapolation for $F_{\rm{D_s}}$.}
  \centerline{\epsfig{file=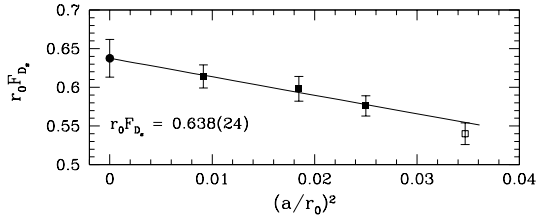,width=\linewidth}}
\label{context}
\vspace{-1.0cm}
\end{figure}

Fig. (\ref{context}) shows the corresponding plateau averages for all simulated values of $\beta$. We excluded the coarsest lattice in the continuum extrapolation which was done linear in $(a/r_0)^2$ because of $O(a)$-improvement. With $b_{\rm A}$ taken from \cite{Bhattacharya:2001ks} we finally quote $r_0F_{\rm{D_s}}=0.638(24)$ as the main result. We also used 1-loop perturbation theory for $b_{\rm A}$ \cite{Sint:1997jx}, since taking $b_{\rm A}$ from \cite{Bhattacharya:2001ks} involves an extrapolation of the data. One then gets $r_0F_{\rm{D_s}}=0.631(24)$. 
Excluding the two coarsest lattices instead leads to $r_0F_{\rm{D_s}}=0.630(34)$.  
Finally, using $r_0=0.5\,$fm, our main result in physical units is $F_{\rm{D_s}}=252(9)\,$MeV.
When using the nucleon mass to set the scale, which corresponds to taking $r_0=0.55\,$fm, we found that $F_{\rm{D_s}}$ decreases by $20\,$MeV \cite{Juttner:2003ns}.
\section{Conclusion}
We present a direct calculation of the heavy-light decay constant $F_{\rm{D_s}}$ with the final result $F_{\rm{D_s}}=252(9)\,$MeV. The error matches that of future experiments, e.g. CLEO-c. We aim at extending the scaling study in the future and these calculations are under way.
We supplement this analysis with more data around the charm mass \cite{Juritalk}. We estimate that $F_{\rm{D_s}}$ decreases by by $20\,$MeV under a scale shift of 10\%.

\end{document}